\documentclass[10pt,sigconf,letterpaper,nonacm]{acmart}

\usepackage{balance}
\usepackage{xspace}
\usepackage{xcolor} 
\usepackage{soul} % need to bar text 

\newif\ifdiff
\difffalse
\ifdiff
    \newcommand{\del}[1]{{{\color{red}\st{#1}}}} 
    \newcommand{\new}[1]{{\color{blue}#1}} 
\else
    \newcommand\del[1]{}
    \newcommand{\new}[1]{{\color{black}#1}}
\fi

\newcommand{\tool}{{SONIC}\xspace}

\newcommand{\eg}{{e.g.,}\xspace}
\newcommand{\ie}{{\it i.e.,}\xspace}

\newcommand{\pb}[1]{\vspace{0.45ex}\noindent{\textbf{\emph{#1}\hspace*{.3em}}}}

\AtBeginDocument{%
  \providecommand\BibTeX{{%
    \normalfont B\kern-0.5em{\scshape i\kern-0.25em b}\kern-0.8em\TeX}}}

\setcopyright{acmlicensed}
\copyrightyear{2018}
\acmYear{2018}
\acmDOI{XXXXXXX.XXXXXXX}

\acmConference[Conference acronym 'XX]{Make sure to enter the correct
  conference title from your rights confirmation emai}{June 03--05,
  2018}{Woodstock, NY}
\acmISBN{978-1-4503-XXXX-X/18/06}

\begin{document}

%\title{\textit{SONIC}: Connecting the Unconnected Using FM Radio and SMS}
\title{\textit{SONIC}: Connect the Unconnected via FM Radio \& SMS}


\author{Ayush Pandey}
\email{ayush.pandey@nyu.edu}
\affiliation{%
  \institution{New York University Abu Dhabi}
  \city{Abu Dhabi}
  \country{United Arab Emirates}
}

\author{Rohail Asim}
\email{rohail.asim@nyu.edu}
\affiliation{%
  \institution{New York University Abu Dhabi}
  \city{Abu Dhabi}
  \country{United Arab Emirates}
}

\author{Khalid Mengal}
\email{kqm1@nyu.edu}
\affiliation{%
  \institution{New York University Abu Dhabi}
  \city{Abu Dhabi}
  \country{United Arab Emirates}
}

\author{Matteo Varvello}
\email{matteo.varvello@nokia.com}
\affiliation{%
  \institution{Nokia Bell Labs}
  \city{New Jersey}
  \country{United States}
}

\author{Yasir Zaki}
\email{yasir.zaki}
\affiliation{%
  \institution{New York University Abu Dhabi}
  \city{Abu Dhabi}
  \country{United Arab Emirates}
}

\renewcommand{\shortauthors}{Trovato and Tobin, et al.}

\begin{abstract}
  As of 2022, about 2.78 billion people in developing countries do not have access to the Internet. Lack of Internet access hinders economic growth, educational opportunities, and access to information and services. Recent initiatives to ``connect the unconnected'' have either failed (project Loon and Aquila) or are characterized by exorbitant costs (Starlink and similar), which are unsustainable for users in developing regions. This paper proposes \tool, a novel connectivity solution that repurposes a widespread communication infrastructure (AM/FM radio) to deliver access to pre-rendered webpages. Our rationale is threefold: 1) the radio network is widely \del{available}\new{accessible} -- currently reaching 70\% of the world  -- even in developing countries, 2) \del{high availability of} unused frequencies \new{are highly available}, 3) while data over sound can be slow, when combined with the radio network it takes advantage of its broadcast nature, efficiently reaching a large number of users. We have designed and built a proof of concept of \tool which shows encouraging initial results.
\end{abstract}

\maketitle

% paper sections 
%%%%%%%%%%%%%%%%%%%%%%
\section{Introduction}
\label{sec:intro}
%%%%%%%%%%%%%%%%%%%%%%
Access to the Internet is vital in today's interconnected world. According to the International Telecommunication Union (ITU) about 2.9 billion people are still offline~\cite{UN_offline} (96\% of them are from developing regions). Several initiatives have attempted to extend Internet connectivity to remote areas, such as Google Loon~\cite{googleLoon} (via high-altitude balloons) and Facebook Aquila~\cite{facebookAquila} (via solar-powered drones). Both projects were eventually halted due to high cost and performance issues (notably Aquila crash~\cite{aquilaCrash}). Starlink~\cite{starlink}, or Internet access provided via low-earth orbit satellites, is another recent solution that is vastly growing (1.5 million users as of June 2023) but characterized by excessive costs for most people from developing regions. 

The goal of this work is to design a novel technology to provide some form of \textit{free} Internet access in developing regions. Our rationale is to leverage existing infrastructures which have broad adoption but were not intended for Internet data dissemination. We propose \textit{\tool}, a system which exploits the FM radio  infrastructure -- currently covering roughly 70\% of the world according to the United Nations~\cite{radioUN} -- to enable the \textit{download} of simplified webpages encoded over sounds.  \tool further relies on the SMS (Short Message Service) network, when available, as uplink to interact with a webpage or request a new one. \tool complements existing initiatives aimed at improving Internet availability. For instance, (expensive) satellite communication can be used to provide Internet connectivity to a few radio stations or towers. These stations can then leverage \tool  to extend access to end-users, making Internet connectivity more affordable and widespread.

\begin{figure}[t]
    \center
    \includegraphics[width = 1\linewidth,trim = 1mm 34mm 1mm 40mm, clip=true]{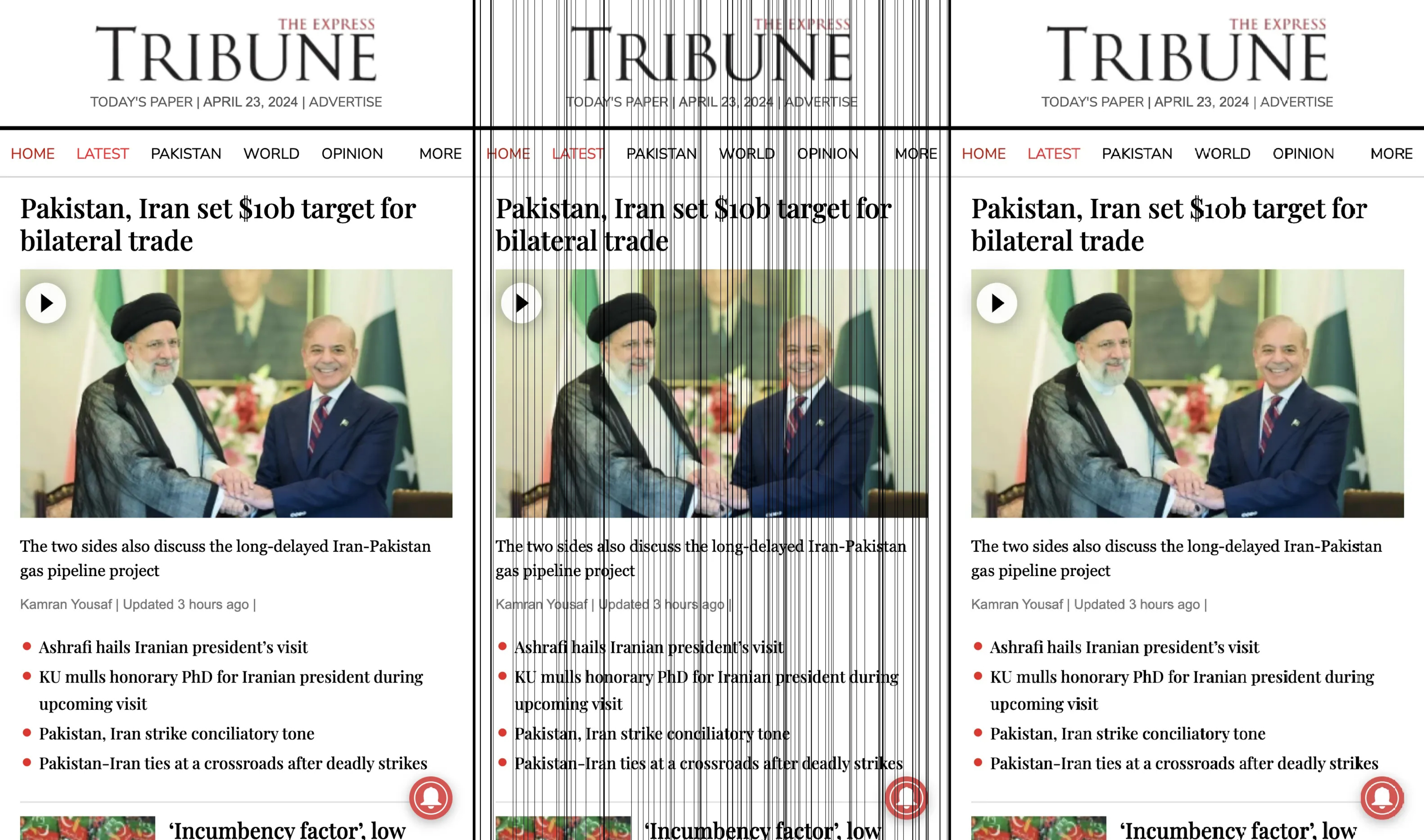}
    \caption{Pre-rendered webpage delivered as \texttt{WebP} via \tool with no frame lost (left), 10\% losses (center), and pixel interpolated on 10\% losses (right).}
    \label{fig:example}
    % \vspace{-0.2in}
\end{figure}

\tool consists of a \textit{client} and a \textit{server}. The server extends an FM transmitter with the capability to load webpages, render them, and then encode their appearances over sounds which are transmitted along with regular radio programs. The client consists of a low-end mobile device and an FM receiver, often integrated. The client is equipped with an application to decode a webpage from the sound received. \tool transmits images of rendered webpages for data savings (about 10x compression) and loss resilience, different from Web objects like HTML and JavaScript. SONIC further uses nearest neighbor pixel interpolation~\cite{nearest-neighbor-interpolation} to recover from errors. Figure~\ref{fig:example} shows that a webpage is still readable despite about 10\% loss rate.

%e.g.,
%{\it i.e.,}
We have built a proof of concept of \tool consisting of a Raspberry Pi acting as FM radio transmitter~\cite{githubFM}, and a Xiaomi Redmi Go~\cite{redmi} acting as a client using its internal FM tuner as well as an external receiver.  For encoding and decoding images of webpages, we rely on the Quiet library~\cite{quiet} and a chunking mechanism we have developed. Using this setup, we demonstrate that a rate of 10kbps is sustainable with no losses when the FM receiver is integrated in the mobile device, and a Received Signal Strength Indication (RSSI) of -65 to -85 dB is achieved. Next, we demonstrate that this transmission rate is sufficient to broadcast 100 popular Pakistani websites with acceptable readability, as confirmed by a user study involving 151 participants. Multiple frequencies can be used to increase the rate, allowing to either broadcast more content or support user requests via uplink.  
%%%%%%%%%%%%%%%%%%%%%%%%%%%%%%%%%%%%%%%
\section{Background and Related Work}
\label{sec:related}
%%%%%%%%%%%%%%%%%%%%%%%%%%%%%%%%%%%%%%
\pb{Data Over Sound:} Several research papers~\cite{ka2016near,lee2015chirp,santagati2016software,roy2017backdoor,bai2020batcomm} and open source tools~\cite{AudioQR,GGwave,quiet} have explored how to transmit data over both sound and ultrasound or inaudible audio frequencies, \ie above 18kHz. Original studies~\cite{ka2016near,lee2015chirp,santagati2016software} focus on the near-ultrasound band (\ie 17-22kHz) and achieve overall low throughput at variable distances: 15bps (up to 25 meters)~\cite{lee2015chirp}, 16bps (2.7 meters)~\cite{ka2016near}, and 2.76kbps (1 meter)~\cite{santagati2016software}.  BackDoor~\cite{roy2017backdoor} achieves higher speed (4kbps and up to 1 meter) by  leveraging the non-linearity in microphone hardware to create a ``shadow'' in the audible frequency range.  BatComm~\cite{bai2020batcomm} further expands on the BackDoor idea by using orthogonal frequency division multiplexing (OFDM~\cite{li2006orthogonal}) to transmit data over multiple orthogonal channels with an ultrasound frequency carrier. Batcomm demonstrates an overall much higher throughput than BackDoor: 47kbps (at 10 centimeters) and 17kbps (at 2.3 meters).  

Among open source tools, AudioQR~\cite{AudioQR} works in the near-ultrasonic frequency band (17.5-19.5kHz) and can reach low speeds of about 100 bps while supporting long distances (up to 150 meters). GGwave~\cite{GGwave} is a tiny data-over-sound library implementing a transmission protocol based on frequency-shift keying (FSK~\cite{watson1980fsk}) which can reach up to 128 bps over short distances. Finally, Quiet~\cite{quiet} enables data over both audible and inaudible frequencies, as well as different modulations such as 1024-QAM~\cite{webb1994modern} and gmsk~\cite{doelz1961minimum}. Quiet claims audible transfers at 7~kbps over the air, and up to 64kbps in cases where two devices are connected over an audio jack cable. 

BatComm is an interesting approach for \tool, but it is  not open source and could not be shared by the authors upon our request. AudioQR is a solid product but not a good fit for \tool since it sacrifices transmission speed for high distance, while we target very low ``air distance'', \eg some phones directly embed an FM receiver (Figure~\ref{fig:flow}). With this respect, GGwave and Quiet are better candidates for \tool; we choose Quiet due to its  higher performance and flexibility. 

\begin{figure}[t]
    \center
    \includegraphics[width = 0.8\linewidth,trim = 0mm 2mm 0mm 0mm, clip=true]{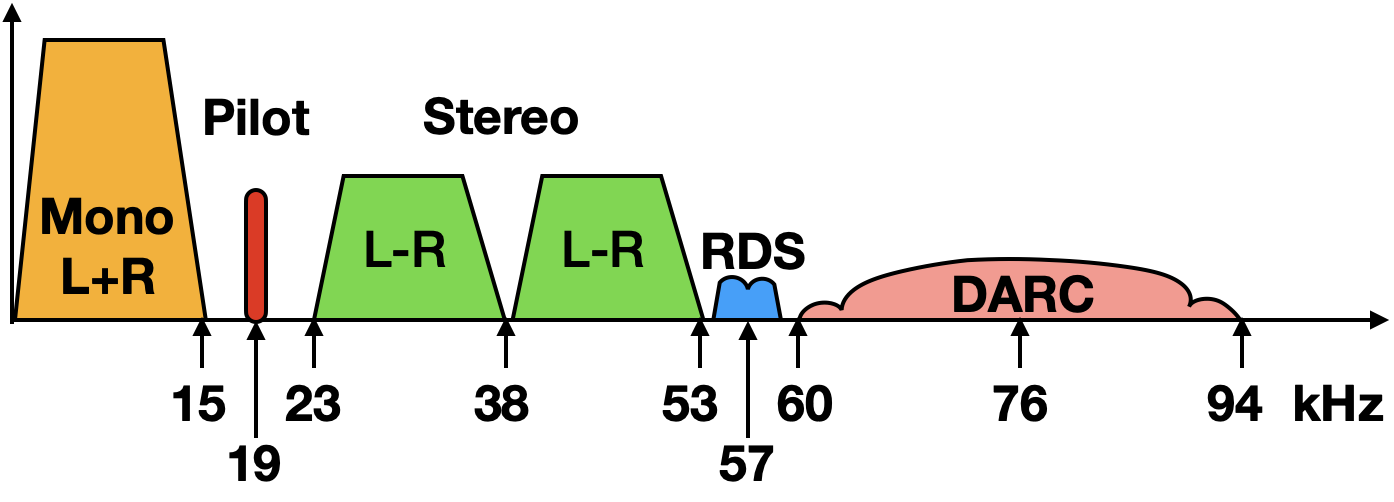}
    % \vspace{-0.1in}
    \caption{Visualization of the FM spectrum.}
    \label{fig:channels}
    % \vspace{-0.2in}
\end{figure}

\pb{Innovative Usage of FM Radio:} To the best of our knowledge, \tool is the first system which uses FM radio to disseminate Web content. Further, \tool combines the FM radio with the SMS infrastructure to provide both a downlink and an uplink. Few other systems have proposed innovative  usage of FM radio as a downlink. RevCast~\cite{schulman2014revcast} leverages the broadcast nature of FM radio for  certificate revocation, \ie the act of invalidating a TLS/SSL certificate before the validity period expires. RevCast relies on Radio Data System (RDS or RDBS in the US), a communications protocol for embedding small amounts of digital information in  FM radio broadcasts. Both RDS and RBDS carry data at 1,187 bits per second on a 57~kHz subcarrier, as shown in Figure~\ref{fig:channels}. The authors show, using a 3~kW commercial radio station and an RDS-enabled smartphone, that RevCast can deliver complete and timely revocation information, anonymously. 

\cite{bozomitu2023drivers, bozomitu2022robust} use FM radio broadcasting to disseminate warning information to drivers. This solution leverages DSSS-CDMA transmissions~\cite{torrieri2005principles}, where each vehicle's geographical position is used as a CDMA key to decode the relevant image. The authors validate their solution using a combination of simulation (MATLAB) and experimental setup (USRP devices~\cite{USRP}) and show data rates of up to 35~kbps. Finally, in 2003 Microsoft used FM subcarrier signals to turn ordinary gadgets into smart gadgets. MSN Direct~\cite{MSNDirect} was a subscription network that sent short text updates over DirectBand (see Figure~\ref{fig:channels}), \ie the 67.65~kHz subcarrier leased by Microsoft from commercial radio broadcasters. This subcarrier delivers about 12~kbit/s of data per tower, for over 100~MB per day per city. MSN Direct has found some success in delivering traffic notifications and weather forecasts to Garmin GPS devices. However, it mainly suffered from: 1) competition with cellular broadband; 2) downlink only communication. Microsoft ended the service in 2012.
%%%%%%%%%%%%%%%%%%%%%
\section{\tool Design}
\label{sec:design}
%%%%%%%%%%%%%%%%%%%%%%%%%%
This Section designs \tool (SOund NIC, or network interface controller), a technology which relies on  FM broadcasting to provide access to simplified webpages encoded over sounds.  The remainder of this section is organized as follows. We first provide an overview of \tool. Next, we dive into the design of its key components.

%%%%%%%%%%%%%%%%%%%%%%%
\subsection{Workflow} 
\label{sec:design:workflow}
\begin{figure}[t]
    \center
    \includegraphics[width = 1\linewidth,trim = 0mm 2mm 0mm 0mm, clip=true]{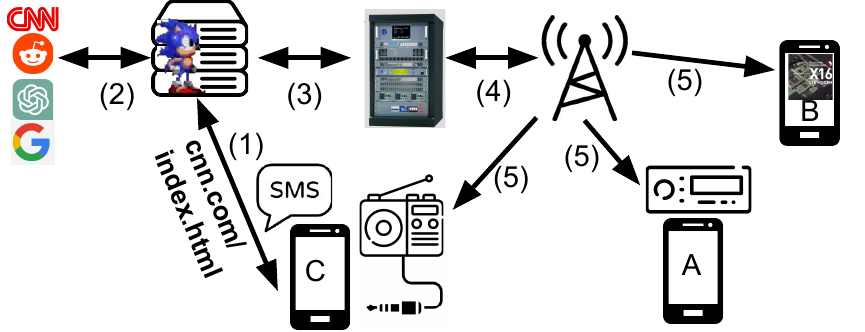}
    \caption{User-C requests a webpage via SMS (1). The SONIC server fetches and renders it into a WebP image (2), then passes it to an FM transmitter (3) for broadcast via sound (4). The webpage is delivered to user-C and any other device currently \textit{listening} to the radio (5).}
    \label{fig:flow}
    % \vspace{-0.2in}
\end{figure}

%%%%%%%%%%%%%%%%%%%%%%%
Figure~\ref{fig:flow} visualizes how \tool works. Each mobile phone showcases a potential \tool user with different capabilities. User-B owns a smartphone equipped with a Qualcomm chip mounting an FM receiver, which is popular in developing regions~\cite{snapdragon}. User-A owns a smartphone with no usable FM receiver, but (s)he is in the proximity of a classic radio, \eg at home or in the car. Both users can receive \tool data, \ie simplified webpages encoded over sounds, but cannot request a specific webpage due to their lack of SMS service. 

User-C connects her phone to an FM radio via the audio jack for higher performance (see Section~\ref{sec:related}) and also pays for Short Message Service (SMS). User-C can both \textit{receive} and \textit{request} webpages. In the figure, user-C is interested in news from \texttt{cnn.com}. This is achieved via an SMS to a \tool provided number (1) which  triggers the generation of a simplified webpage for \texttt{cnn.com} (2), \eg the landing page, and its broadcasting over the FM radio (3)(4), so that user-C (as well as user-A and user-B) will eventually receive it (5).

\vspace{0.02in}
\pb{\tool Server:} It is responsible for generating simplified webpages to broadcast via FM radio. We assume that the FM radio infrastructure consists of multiple transmitters (and frequencies) at different locations, with some form of Internet access. Accordingly, the FM transmitters can receive simplified webpages to be encoded via sound, and then transmit them. A central \tool server is responsible to produce such webpages and inform the respective transmitters.  

The \tool server produces simplified webpages according to user requests. Further, it maintains a list of the most popular websites in a region that are preemptively \textit{pushed} to users in an attempt to improve their experience. For example, popular news sites can be pushed early in the morning. 

\vspace{0.02in}
\pb{\tool Client:} As visualized in Figure~\ref{fig:flow}, users access \tool using different hardware configurations. The minimum configuration implies a smartphone and access to an FM radio, either over the air or via an audio jack, which provides the \textit{downlink}. An SMS service is also needed for users interested in having access to the \textit{uplink}, \eg request a specific webpage or query a search engine. Software-wise, \tool runs as a standalone user space application we describe next. 

The \tool application allows to \textit{browse} webpages previously received via the FM radio and request new ones. Accordingly, the app shows a catalog of available webpages, organized by content, popularity, and/or user interest.  Within a webpage, a user might be interested in visiting some internal pages by following classic hyperlinks. If the requested internal page is locally available, \ie it was previously received via the FM radio, then the page would instantly load. If not, an active uplink is required to request such a page. Similarly, \tool users with an active uplink can request additional  webpages not previously received, as well as send queries to search engines (\eg Google and Duckduckgo) and AI chatbots (\eg chatGPT).  Note that pages requiring authentication, such as online banking and social media, are not supported yet due to their privacy risk, as discussed in Section~\ref{sec:design:discussion}. 
\tool clients request \textit{new} webpages, \ie not previously received and locally cached, via SMS. Each request contains the URL, \eg \texttt{cnn.com/index.html}, and the geographic location of the user. The latter is needed by \tool server to inform the proper transmitter along with its frequency, as discussed below. As the request is received, the \tool server produces a \textit{simplified} version of the webpage, either from its cache, \eg if recently requested by another user, or by directly accessing it. The simplified page is then transferred to the radio transmitter, which can physically reach the current user's location.  Meanwhile, \tool server quickly responds to the user via SMS to acknowledge the request, and provide an estimate on when the page will be received. When the webpage is received, it is inserted in a cache with expiration date set according to a time indicated by the server. 

%%%%%%%%%%%%%%%%%%%%%%%%%%%%%%%%%%%%%%%%%
\subsection{Simplified Webpage Creation}
\label{sec:design:webpage}
%%%%%%%%%%%%%%%%%%%%%%%%%%%%%%%%%%%%%%%%%
According to~\cite{webAlmanac}, the average webpage size is around 2~MB for mobile sites. Given the low speeds of data over sound, \ie tens of kbps at best (see Section~\ref{sec:related}), broadcasting 2~MB via \tool might require tens of minutes. Further, receivers might not be able to reconstruct the received page, given that most Web components, like JavaScript, would break in presence of unrecoverable errors. 

It follows that we need a mechanism to: 1) greatly compress webpages, 2) make webpages resilient to noise. Many solutions exist to reduce webpage sizes, such as compression proxies~\cite{amp, webLight}, reader modes~\cite{ghasemisharif2019speedreader,safari}, JavaScript cleaners~\cite{JSCleaner,slimweb,jsanalyzer}
and redundant code removal~\cite{kupoluyi2022muzeel,malavolta2023javascript}. However, these solutions are still affected by noise and would thus require a large amount of FEC, which implies dimensioning the system for a worst-case receiver. Instead, we would like a solution whose performance decreases as a function of the receiver's RSSI, similarly to how the audio quality is reduced for users with bad reception. 

We took inspiration from the work in~\cite{bozomitu2023drivers, bozomitu2022robust}, which shows how the quality of images transmitted over RDS decreases as a function of a receiver's RSSI. Accordingly, we propose to transmit \textit{images} of loaded webpages instead of actual Web files (html, js, css, etc.). Such images achieve both our goals since they compress a page, \eg a 2~MB page can be compressed as a few hundred KB image, and they are loss-resilient, as despite missing pixels, users might still understand the content provided. 

DRIVESHAFT (DS)~\cite{bhardwaj2018driveshaft} is a system to speed up the mobile Web by distributing crowdsourced screenshots of webpages. While the overall system is quite different, \eg it involves a complex mechanism to merge harvested screenshots while stripping them of private information, \tool could leverage some of DS's components, \eg how to add interactivity to the screenshots.  Unfortunately, DS's code is not available, and we could not experiment with and integrate it with \tool. We here discuss \tool's mechanism to form images of webpages and its relationship with DS. 

\begin{figure*}[tb]
    \includegraphics[width=\linewidth]{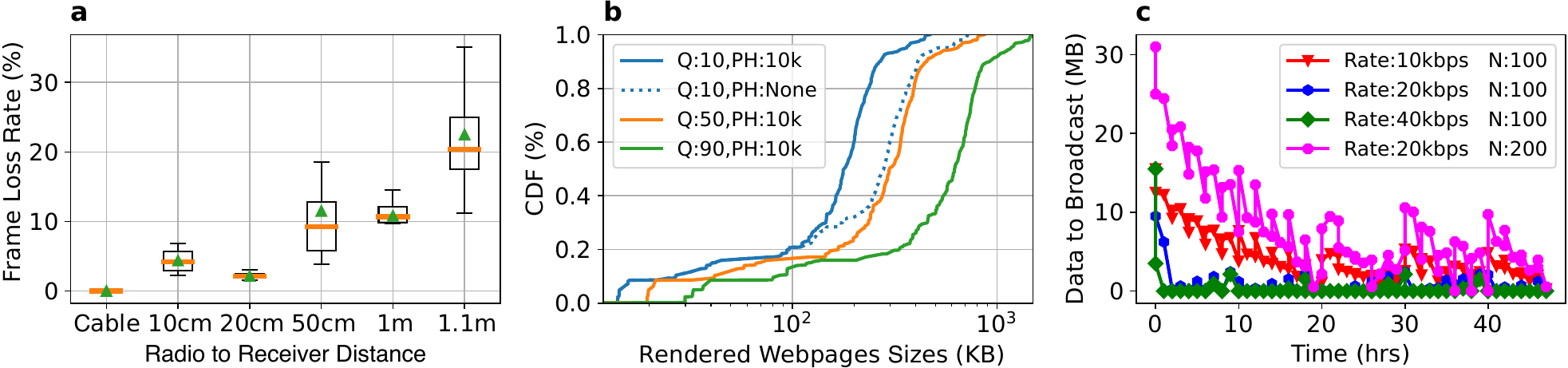}  
    \caption{Preliminary performance evaluation of \tool. (a) Frame loss rate as a function of the over-the-air distance between FM receiver and a \tool client. ``Cable'' refers to a distance of zero, \ie either an integrated FM receiver on the phone or a jack audio cable connection. (b) CDF of the size of images (\texttt{WebP}) of rendered webpages, assuming variable image quality (Q) and pixel height (PH). (c) Evolution over time of the amount of data to be broadcasted as a function of transmission rates and number of webpages (N).}
    % \vspace{-0.15in}
    \label{fig:benchmarking}
\end{figure*}

\vspace{0.02in}
\pb{Interactivity:} modern webpages allow user interaction either via simple hyperlinks or with menus, search boxes, etc. Images are instead static by definition. To allow some interactivity, ~\cite{bhardwaj2018driveshaft} proposes to build \textit{click maps}, \ie maps containing a list of <x,y> coordinates of where an image is interactive. We adopt the same approach: as the user clicks on such coordinates, \tool informs the server (via SMS, if available) and requests the next image, \eg an internal page accessed via a hyperlink, unless it is already available in the cache. Due to the more challenging network conditions of \tool, \eg potentially seconds in uplink and minutes in downlink, we limit the interactivity to hyperlinks only. 

\vspace{0.02in}
\pb{Images Format:} DS requires lossless \texttt{png} due to its need to desensitize and merge crowdsourced screenshots. This is not a requirement for \tool; we thus resort to \texttt{WebP}~\cite{webp}, a modern image format that provides superior lossless and lossy compression. We capture images of webpages as \texttt{WebP} with 10\% quality, which significantly reduces the size while remaining acceptable, as we evaluate in Section~\ref{sec:eval}. These images are created with 1,080 pixels width, and a maximum of 10k pixels height. The idea is to allow a user to ``scroll'' down on an image, as commonly done on a webpage, while avoiding to waste broadcasted data, given that user interest is lower for content below the fold. Depending on the mobile phone screen resolution, and using the scaling factor (\ie mobile phone screen width / 1,080), the images are resized by multiplying both the width and height with the scaling factor. The scaling factor is also used to scale the coordinates of the click maps. 

%%%%%%%%%%%%%%%%%%%%%%%%%%%%%%%%%%%%%%%%%%%
\subsection{Sonic Data Transmission}
\label{sec:design:sonic}
%%%%%%%%%%%%%%%%%%%%%%%%%%%%%%%%%%%%%%%%%
Using the Quiet library~\cite{quiet}, we create a new transmission profile inspired by their ``audible-7k-channel''. The new profile uses Orthogonal Orthogonal Frequency-Division Multiplexing (OFDM) -- a multi-carrier modulation system where data is transmitted as a combination of orthogonal narrow-band signals known as sub-carriers -- with 92 sub-carriers. The data rates achieved by this profile reach 10~kbps. Upon transmitting a rendered page, we first divide the image vertically into multiple partitions, each with a width of 1 pixel. Each partition is then divided into fixed-sized frames of 100 bytes each. Each frame carries a partition and a sequence number used to reassemble the image on the receiver end. We further leverage the error detection and correction schemes using: crc32 as the checksum, the inner Forward Error Correction (FEC) scheme (v29), and an outer FEC scheme (rs8). These allow the receiver to correct transmission errors.

In the case of lost frames, the \tool receiver uses nearest neighbor pixel interpolation~\cite{nearest-neighbor-interpolation} to correct missing pixels. Accordingly, missing pixels are replaced with the value of their adjacent pixel, prioritizing the left pixel given that the webpage consists mostly of text read from left to right. In literature, more advanced and better performing techniques exist to recover missing pixels in images~\cite{gradient-compressive-sensing, deep-learning-data-reconstruction, deep-image-inpainting}, leveraging deep neural networks to learn the patterns and structures of the image or leveraging sparsity and gradients in the data to fill in the missing regions. These techniques are both memory- and CPU-intensive, far beyond what a low-end mobile device can support today.

%%%%%%%%%%%%%%%%%%%%%%%%
\subsection{Discussion}
\label{sec:design:discussion}
%%%%%%%%%%%%%%%%%%%%%%%%
\pb{Incentives and Monetization:} For users, the incentive is clear since they get access to a simplified Web in areas where such content is currently inaccessible. For \tool providers, they could directly charge their users. This can be challenging since, for example, downlink users are completely passive leaving no trace to when any content is accessed. Alternatively, they could bundle their service with an SMS service, thus offering a paid service for users who can request content in real time, while being \textit{free} for other users. Anyway, the cost of broadcasting over the radio is constant regardless of the number of ``listeners''. 

Finally, a more interesting monetization scheme arises when considering the current business model of radio stations: increasing their user base to sell more advertisements. \tool represents a novel service that radio stations can offer to attract more users, thus increasing their revenues via ads. Further, ads can now also be integrated with webpages, expanding from audio only to images. 

\pb{Privacy Concerns:} In essence, \tool functions similarly to Web acceleration tools like Google AMP~\cite{amp} and WebLight~\cite{webLight} which rewrite the content before sending it to a user. To do so, \tool requires full visibility on the requested URL, as well as the content being currently accessed by a user. It follows that a \tool server could collect enough information to build user profiles and eventually violate user privacy. Differently from these solutions, \tool relies on a broadcast medium (the FM radio), thus not knowing precisely how many users might be accessing a given content. Indeed, no privacy violation is possible for downlink-only users who can never request any content but just occasionally browse popular content requested by other users in their area. 

\pb{Content Limitations:} \tool does not provide access to webpages behind a login, such as a personal bank website or a social media account. Such access is impossible for downlink-only users; for users with an uplink, they would have to share their username and password to allow the SONIC server to retrieve the required content. Further, the resulting page (\eg showing the balance of the user) would be broadcasted to potentially many users, adding to the privacy issue. 

Next, \tool lacks support for video, which is used at large in today's Web not only by video streaming services but also by news websites, social media, etc. Simply put, \tool does not have enough bandwidth to support video streaming; instead, videos are replaced by their thumbnails which are not ``clickable''. Similarly, \tool transmits simplified (pre-rendered) webpages with no support of complex visual effects normally achieved via either JS or CSS.

\pb{Scalability:} \tool is a best effort system, aiming to offer few webpages to a large number of users -- anyone in the range of an FM radio -- despite some very small bandwidth. This is achieved by bending the current strictness of webpages, and accepting some visual quality degradation as well as lack of full interactivity. Further, the catalog of webpages to be served has to be small (see Figure\ref{fig:benchmarking}(c)) and thus properly curated to offer interesting and useful content to its users.
% \vspace{-5pt}
%%%%%%%%%%%%%%%%%%%%%%%%%%%%%%%%%%%%
\section{Preliminary Evaluation}
\label{sec:eval}
%%%%%%%%%%%%%%%%%%%%%%%%%%%%%%%%%%%%

\pb{Methodology:}  We build an FM transmitter using a Raspberry Pi, a 20cm wire acting as an antenna connected to GPIO pin 4 of the Pi, and the code from~\cite{githubFM}, which uses the general clock output to produce frequency modulated radio communication in the range from 1MHz to 250MHz. We use 93.7MHz since it is unused at our location. For data transmission, we use the Mono channel of the FM baseband (30Hz to 15KHz, see Figure~\ref{fig:channels}) with a carrier center frequency of 9.2KHz. We envision that other bands can be used to increase the data rate, \eg using the left and right band of the Stereo channel, or even the DARC band. We left this exploration as future work. As FM receivers, we consider the scenario showcased in Figure~\ref{fig:flow}. User-B leverages the internal FM tuner of a Xiaomi Redmi Go~\cite{redmigo}, and a pair of headphones as an antenna. User-A is another Redmi Go receiving sound from a nearby FM radio over the air. Finally, User-C is a Redmi Go connected to the same FM radio but using the audio jack. We ensure high RSSI between transmitter and receiver.

For encoding data over sound, we use Quiet~\cite{quiet} as discussed in Section~\ref{sec:design:sonic}. For decoding, we develop an Android application extending the sample application~\cite{quietAndroid} from Quiet. We approximate user-B using the NextRadio app~\cite{nextRadio} to receive the FM radio signal and a jack cable to act both as an antenna and to share the received sound to a second device which runs the \tool app. In the future, we will integrate FM radio receiving within the \tool app. 

For the content to be transmitted, we assume a deployment in Pakistan. Accordingly, we selected the 25 most popular Pakistani websites from the Tranco list~\cite{tranco} filtered using the \texttt{.pk} domain name. For each landing page, we select three random internal pages, resulting in a total of 100 webpages (25 landing, and 75 internal). We rendered these 100 pages in Chrome on hourly basis for three days, while generating variable quality images of the rendered webpages. 

\vspace{0.1in}
\pb{General Results:}  Figure\ref{fig:benchmarking}(a) shows the frame loss rate, at the receiver as a function of the distance between the phone and the receiving radio. ``Cable'' refers to a scenario where the receiver is directly connected to the FM radio, either using the internal FM tuner (User-B in Figure~\ref{fig:flow}) or an actual cable (User-C in Figure~\ref{fig:flow}). Each experiment is repeated 10 times, and we assume high RSSI (-70 db or higher) at the FM receiver. The figure shows no frame loss recorded over cable, and up to 10-20\% frame losses (at the median) when considering about one meter of over-the-air communication between a SONIC-enabled phone and FM receiver. The loss rate growth is not completely linear with the increased air distance, as the alignment between speaker and microphone also has a significant impact. We did not control for this variable, as \tool users would likely not do so as well. We also observe a 100\% loss rate at distances above 1.1m.

Figure~\ref{fig:example} visualizes the impact of 10\% losses. Overall, the impact is significant but \textit{tolerable}, unless unlucky scenarios where losses are concentrated in important parts of the webpage, like title and text. In our early prototype, each portion of an image is transmitted equally; one optimization consists of adopting a dynamic scheme with higher error protection for important parts of an image/webpage. Nevertheless, we envision that \tool users would mostly rely on their phone's internal FM tuner -- or a cable connection to a radio -- to improve their performance and minimize lost frames. 

Figure~\ref{fig:benchmarking}(b) shows the Cumulative  Distribution Function (CDF) of the size of images (\texttt{WebP}) of rendered webpages assuming variable image quality (Q). \texttt{WebP} image quality is defined on a scale from 0 (worst) to 95 (best). The figure also reports the pixels height (PH), either ``none'' (\ie maximum length of the webpage loaded) or cropped at 10K pixels (see Section~\ref{sec:design:webpage}). Assuming a low quality of 10, most webpages can be compressed to less than 200~KB, which instead requires 700~KB at a high quality of 90. For \tool, given the challenging network conditions of the FM radio, we opt for a low quality of 10 given that no significant image degradation is observed. The figure also shows that cropping to 10k pixels saves about 100~KB for 75\% of the webpages. The tails of the CDFs show images with twice the size of the 90th percentile, \eg 500~KB versus 250~KB when considering Q equal to 10 and PH equal to 10~k. These images can take significant time to be delivered, \eg up to 6/7 minutes if assuming a single delivery frequency and a 10~kbps rate (see Section~\ref{sec:design:sonic}), and might thus require some further optimizations.

\begin{figure}[t]
    \includegraphics[width=0.9\linewidth]{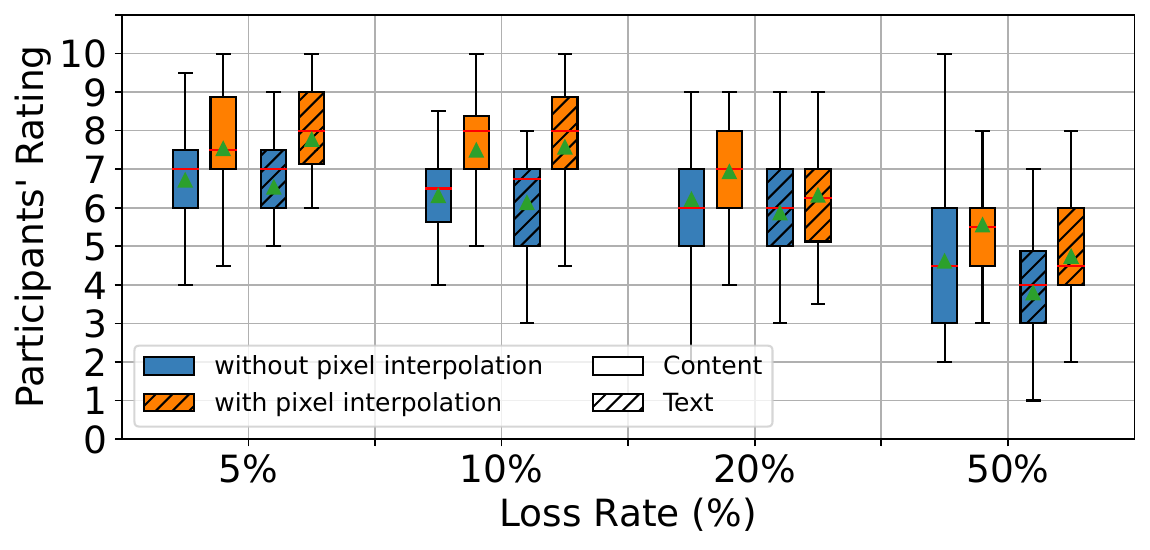}  
    % \vspace{-10pt}
    \caption{Distribution of median user ratings (0-10 Likert scale) per top 50 Pakistani webpages, for question-b (plain) and question-a (hatched), with and without pixel interpolation (orange and blue, respectively).}    
    % \vspace{-0.1in}
\label{fig:lossrateimpact}
\end{figure}

Finally, Figure~\ref{fig:benchmarking}(c) shows the evolution over time of the amount of data broadcasted with \tool assuming the above 100 Pakistani webpages, and the current rate we have verified (10kbps) as well as 20 and 40~kbps which can be achieved via multi-frequency (see Section~\ref{sec:design}). 
Note that we truncated the figure at two days -- despite three days of data -- since the pattern repeats and it helps improve its visibility. 

Figure~\ref{fig:benchmarking}(c) shows that, with the current transmission rate (10~kbps), \tool can only work in broadcast mode. In fact, the amount of data to be sent -- due to variation in the supported webpages -- rarely reaches zero (for which a rate of 20 or 40~kbps is required). However, \tool is \textit{scalable}, meaning that the amount of data to be sent does not grow indefinitely. 

\vspace{0.02in}
\pb{Variable RSSI:} 
We evaluate SONIC's frame losses across various levels of RSSI (Received Signal Strength Indication). RSSI measures signal strength in decibels, typically ranging from 0 (strongest) to -120 (lowest). FM requires RSSI comprised between -65 and -80 dB; e.g., radio stations in our area exhibit RSSI between -70 and -80 dB.

We utilize the TR508 FM Transmitter~\cite{retekess}, which supports transmissions up to 1~km. We pair the SONIC client in ``cable'' mode (i.e., with no additional frame loss rate, as shown in Figure \ref{fig:benchmarking}(a)) with the Real FM Radio app~\cite{real_fm_app}, which records RSSI values. We broadcast on an empty FM frequency and manually explore distances between transmitter and receiver to observe RSSI signals within the -65 to -90 dB range. At approximately 5 dB intervals, we transmit a single webpage up to 10 times and measure SONIC’s frame loss rate. For the RSSI range from -65 to -85 dB, we consistently observe no frame losses. For the -85 to -90 dB range, we record a fluctuating frame loss rate between 2 and 15\%. As expected~\cite{sonicwall}, for RSSI below -90 dB, we are unable to receive any frames.

\vspace{0.02in}
\pb{User Study:} To assess the impact of losses on webpage readability, we create screenshots of the top 50 Pakistani webpages (according to Tranco~\cite{tranco}) with synthetic variable losses (5\%, 10\%, 20\%, and 50\%). Missing pixels are either shown as dark or corrected using nearest neighbor pixel interpolation (see Section~\ref{sec:design:sonic}). This results in 400 screenshots (50 pages $\times$ 4 loss rates $\times$ 2 approaches, i.e., with and without pixel interpolation). We then recruit 151 students from a university in Pakistan to evaluate 20 randomly selected screenshots, thus averaging at least 7 ratings per screenshot. 

Students rate screenshots on a Likert scale from 0 to 10 based on two questions: a) \textit{Indicate your perception of content understanding in the above image (0 = extremely unclear, 5 = neutral, and 10 = extremely clear)}; and b) \textit{Rate the readability of the text in the above image considering the level of noise present (0 = very difficult to read, 5 = neither easy nor difficult, and 10 = very easy to read)}. Figure~\ref{fig:lossrateimpact} shows boxplots of the median rating per page for both questions (plain for question-b and hatched for question-a), differentiating whether pixel interpolation is used (orange) or not (blue). Overall, the figure demonstrates the benefits of pixel interpolation to enhance both text readability and content comprehension, improving the rating by at least one point regardless of the loss rate. For example, even at a loss rate of 20\%, observed at a distance of 1.1 meters and above (see Figure~\ref{fig:benchmarking}(a)), pixel interpolation achieves a median content readability score of 7 (somewhat clear understanding). Text readability is more susceptible to losses, but still acceptable even with 20\% losses.

\balance

\bibliographystyle{ACM-Reference-Format}
\bibliography{main}

\end{document}